\documentclass[twocolumn,prb,aps,floatfix,superscriptaddress]{revtex4-1}

\usepackage{color}
\usepackage{bm}
\usepackage{graphicx}
\usepackage{amsmath}
\usepackage{amssymb}
\usepackage{bbold}
\usepackage{setspace}
\usepackage{epstopdf}
\usepackage{placeins}
\usepackage[export]{adjustbox}

\newcommand{\figref}[1]{Fig. \ref{#1}}
\newcommand{\secref}[1]{Sec. \ref{#1}}
\newcommand{\tabref}[1]{Tab. \ref{#1}}
\newcommand{\appref}[1]{Appendix \ref{#1}}

\newcommand*{\bplqt}{%
  \text{
    \includegraphics[
      height=1.2ex,
      valign=M,
      raise=\fontdimen22\textfont2,
    ]{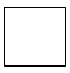} }
}
\newcommand*{\fplqt}{%
  \text{
    \includegraphics[
      height=1.2ex,
      valign=M,
      raise=\fontdimen22\textfont2,
    ]{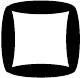} }
}
\newcommand*{\vvemptylink}{%
  \text{
    \includegraphics[
      height=1.8ex,
      valign=M,
      raise=\fontdimen22\textfont2,
    ]{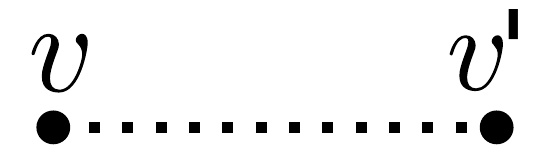} }
}
\newcommand*{\emptylink}{%
  \text{
    \includegraphics[
      height=1.8ex,
      valign=M,
      raise=\fontdimen22\textfont2,
    ]{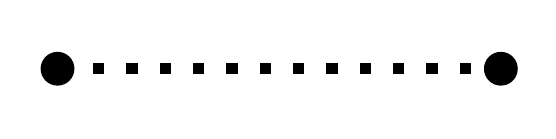} }
}
\newcommand*{\rightarrowlink}{%
  \text{
    \includegraphics[
      height=1.8ex,
      valign=M,
      raise=\fontdimen22\textfont2,
    ]{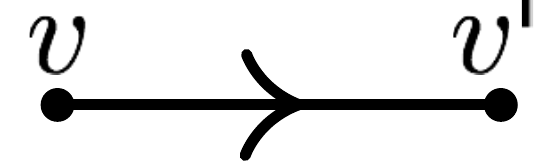} }
}
\newcommand*{\leftarrowlink}{%
  \text{
    \includegraphics[
      height=1.8ex,
      valign=M,
      raise=\fontdimen22\textfont2,
    ]{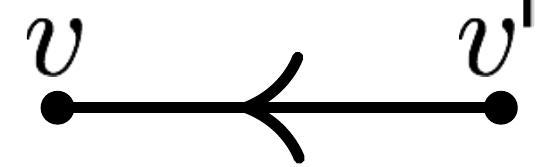} }
}

\newcommand*{\hrzplqt}{%
  \text{
    \includegraphics[
      height=1.5ex,
      valign=M,
      raise=\fontdimen22\textfont2,
    ]{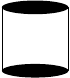} }
}
\newcommand*{\vrtplqt}{%
  \text{
    \includegraphics[
      height=1.5ex,
      valign=M,
      raise=\fontdimen22\textfont2,
    ]{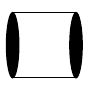} }
}
\newcommand*{\threeleftplqt}{%
  \text{
    \includegraphics[
      height=1.5ex,
      valign=M,
      raise=\fontdimen22\textfont2,
    ]{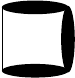} }
}
\newcommand*{\oneleftplqt}{%
  \text{
    \includegraphics[
      height=1.5ex,
      valign=M,
      raise=\fontdimen22\textfont2,
    ]{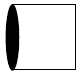} }
}
\newcommand*{\twoupplqt}{%
  \text{
    \includegraphics[
      height=1.5ex,
      valign=M,
      raise=\fontdimen22\textfont2,
    ]{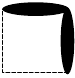} }
}
\newcommand*{\twodownplqt}{%
  \text{
    \includegraphics[
      height=1.5ex,
      valign=M,
      raise=\fontdimen22\textfont2,
    ]{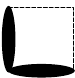} }
}
\newcommand*{\svrtmon}{%
  \text{
    \includegraphics[
      height=3.5ex,
      valign=M,
      raise=\fontdimen22\textfont2,
    ]{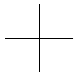} }
}
\newcommand*{\svrtdmr}{%
  \text{
    \includegraphics[
      height=3.5ex,
      valign=M,
      raise=\fontdimen22\textfont2,
    ]{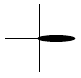} }
}
\newcommand*{\svrttrmr}{%
  \text{
    \includegraphics[
      height=3.5ex,
      valign=M,
      raise=\fontdimen22\textfont2,
    ]{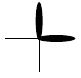} }
}
\newcommand*{\svrttrmrtwo}{%
  \text{
    \includegraphics[
      height=3.5ex,
      valign=M,
      raise=\fontdimen22\textfont2,
    ]{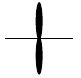} }
}
\newcommand*{\svrtquat}{%
  \text{
    \includegraphics[
      height=3.5ex,
      valign=M,
      raise=\fontdimen22\textfont2,
    ]{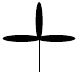} }
}
\newcommand*{\svrtpent}{%
  \text{
    \includegraphics[
      height=3.5ex,
      valign=M,
      raise=\fontdimen22\textfont2,
    ]{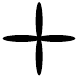} }
}
\newcommand*{\pentmvcreate}{%
  \text{
    \includegraphics[
      height=3.5ex,
      valign=M,
      raise=\fontdimen22\textfont2,
    ]{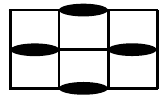} }
}
\newcommand*{\pentmvanal}{%
  \text{
    \includegraphics[
      height=3.5ex,
      valign=M,
      raise=\fontdimen22\textfont2,
    ]{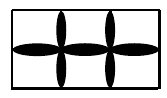} }
}
\newcommand*{\penmvhrzleft}{%
  \text{
    \includegraphics[
      height=3.5ex,
      valign=M,
      raise=\fontdimen22\textfont2,
    ]{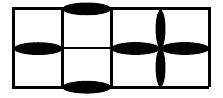} }
}
\newcommand*{\penmvhrzright}{%
  \text{
    \includegraphics[
      height=3.5ex,
      valign=M,
      raise=\fontdimen22\textfont2,
    ]{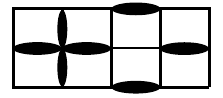} }
}
\newcommand*{\penmvdiagup}{%
  \text{
    \includegraphics[
      height=5.5ex,
      valign=M,
      raise=\fontdimen22\textfont2,
    ]{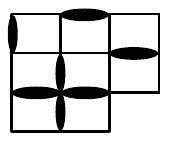} }
}
\newcommand*{\penmvdiagdown}{%
  \text{
    \includegraphics[
      height=5.5ex,
      valign=M,
      raise=\fontdimen22\textfont2,
    ]{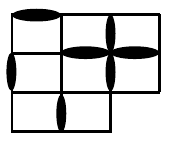} }
}
\newcommand*{\monomermvA}{%
  \text{
    \includegraphics[
      height=3.5ex,
      valign=M,
      raise=\fontdimen22\textfont2,
    ]{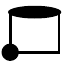} }
}
\newcommand*{\monomermvB}{%
  \text{
    \includegraphics[
      height=3.5ex,
      valign=M,
      raise=\fontdimen22\textfont2,
    ]{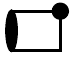} }
}
\newcommand{\ket}[1]{| #1 \rangle}
\newcommand{\eket}[1]{\left | #1 \right \rangle}
\newcommand{\ebra}[1]{\left \langle #1 \right |}
\newcommand{\exval}[1]{\langle #1 \rangle}
\newcommand{\Eqref}[1]{Eq. \eqref{#1}}
\newcommand{\RK}{\mathrm{RK}}
\newcommand{\HGQDM}{H_\mathrm{GQDM}}
\newcommand{\HQDM}{H_\mathrm{QDM}}
\newcommand{\HIGT}{H_\mathrm{IGT}}
\newcommand{\HQDPM}{H_\mathrm{QDPM}}
\newcommand{\Mrot}{M_{\rm{rot}}}
\newcommand{\Mperp}{M_\perp}
\newcommand{\Mpar}{M_{\vert\vert}}
\newcommand{\MstagSq}{M_{\rm{stag}}^2}
\newcommand{\McolSq}{M_{\rm{col}}^2}

\begin{document}

\title{$Z_3$ topological order in the quantum dimer-pentamer model}

\author{Owen Myers}
\email{omyers@uvm.edu}
\affiliation{Department of Physics, University of Vermont, Burlington, VT 05405, USA}

\author{C. M. Herdman}
\affiliation{Institute for Quantum Computing, University of Waterloo, Ontario, N2L 3G1, Canada}
\affiliation{Department of Physics \& Astronomy, University of Waterloo, Ontario, N2L 3G1, Canada}
\affiliation{Department of Chemistry,  University of Waterloo, Ontario, N2L 3G1, Canada}

\begin{abstract}
We introduce the quantum dimer-pentamer model (QDPM) on the square lattice. This model is a generalization of the square lattice quantum dimer model as its configuration space comprises fully-packed hard-core dimer coverings as well as dimer configurations containing pentamers, where four dimers touch a vertex. Thus in the QDPM, the fully-packed, hard-core constraint of the quantum dimer model is relaxed such that the local dimer number at each vertex is fixed modulo 3, resulting in an exact local $Z_3$ gauge symmetry. We construct a local Hamiltonian for which the Rokhsar-Kivelson (RK) equal superposition state is the exact ground state and has a 9-fold topological degeneracy on the torus. Using Monte Carlo calculations, we find no spontaneous symmetry breaking in the RK wavefunction and that its dimer-dimer correlation function decays exponentially. By doping the QDPM RK state with a pair of monomers, we demonstrate that $Z_3$ electric charges are deconfined. Additionally, we introduce a $Z_3$ magnetic string operator that we find decays exponentially and shows no signatures of magnetic vortex condensation and with correlations. These numerical results suggest that the ground state of the QDPM is a dimer liquid with $Z_3$ topological order.
\end{abstract}

\maketitle

\section{Introduction}

Strongly interacting quantum many-body systems are known to support a variety of exotic quantum phases of matter with no classical analogue. Among these are topologically ordered quantum liquids; their non-trivial topological order can not be identified by a local order parameter as these phases of matter don't break conventional symmetries~\cite{Wen1990,Nayak2008}. Both because of of the exotic nature of topological phases and their potential use for quantum information processing~\cite{Freedman2001,Kitaev2003}, there is great interest in the development of theoretical models which display these phases and could provide candidate experimental systems to constructively engineer topological order~\cite{Duan2003,Jaksch2005,Lewenstein2007,Jiang2008d,Weimer2010a,Herdman2010c,Fowler2012a}. 

A variety of exactly soluble lattice models are known to support ground states with topological order that ranges from the simplest variety, $Z_2$ topological order~\cite{Kitaev2003,Wen2003}, to those with non-Abelian anyonic excitations~\cite{Levin2005a,Kitaev2006a}; such models generally involve complex many-body interactions, and thus it is desirable to find simpler models supporting topological order that are closer to what is accessible experimentally. Among the simplest classes of models displaying topological quantum liquid phases are those with local geometric constraints; the constraints inhibit the formation of symmetry breaking local order and given sufficient dynamics, may lead to a symmetric liquid ground state. The quantum dimer model (QDM) describing dimer degrees of freedom living on the links of a lattice is one of the most basic locally constrained models~\cite{Rokhsar1988,Moessner2008}; the hard-core, fully-packed QDM has a local constraint that requires that exactly one dimer touches each vertex.

\begin{figure}[t!]
    \centering
    \includegraphics[width=0.9\columnwidth]{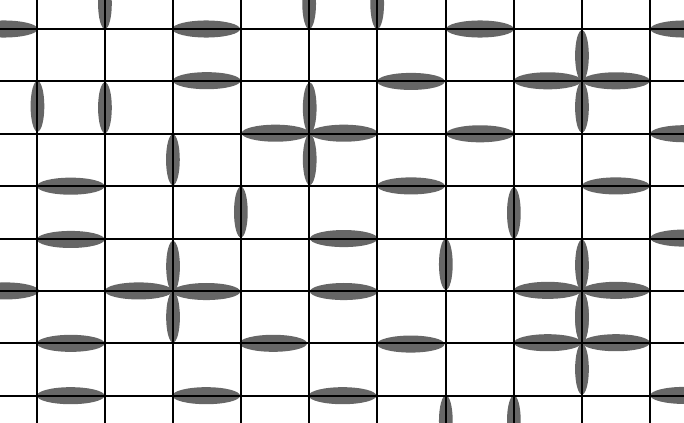}
    \caption{An example of an allowed configuration in the quantum dimer-pentamer model. The local constraint requires that each vertex is touched by either one dimer or a four dimers (a pentamer).}
    \label{fig:QDPMex}
\end{figure}

While on non-bipartite two-dimensionsal lattices, the QDM has a topologically ordered ground state~\cite{Moessner2001a,Fendley2002}, on the square lattice there is only a symmetric liquid ground state at an isolated critical quantum point~\cite{Leung1996,Syljuasen2006}. If the hard-core constraint of the square lattice QDM is relaxed such that only the parity of the number of dimer touching each vertex is fixed, the simplest non-trivial dynamics leads to a $Z_2$ topologically ordered ground state~\cite{Kitaev2003,Wen2003}. The existence of topologically ordered ground states in these models is intimately related to the local constraint which imposes a local gauge symmetry in the Hilbert space; the topological order that arises is directly related to a discrete gauge theory with the appropriate local gauge constraint~\cite{Moessner2001}.

Given that these models with dimer degrees of freedom and local constraints posses the simplest non-trivial local Hilbert space at each link, they provide a potentially simpler route to engineering exotic phases. The local constraints can be enforced by a local potential energy penalty and the minimal dynamics often arise at the lowest non-trivial order in perturbation theory from a related spin model\cite{Albuquerque2008}. We will refer to such models with dimer degrees of freedom on links and local constraints as \emph{generalized quantum dimer modes} (GQDMs). Previous work has demonstrated that GQDMs on the square lattice support gapped $Z_2$ topologically ordered phases as well as gapless quantum critical points. On non-bipartite lattices, GQDMs support $Z_2$ topological order~\cite{Moessner2001a,Misguich2002} as well as doubled-semion phases~\cite{Qi2015,Buerschaper2014}. A full characterization of all exotic phases that can exist within this framework is thus desirable.

In this work we extend this paradigm to a new GQDM on the square lattice, what we term the \emph{quantum dimer-pentamer model} (QDPM). In the QDPM, the local constraint is relaxed from that of the QDM to allow either $1$ or $4$ dimers to touch each vertex, as shown in figure \figref{fig:QDPMex}. Correspondingly, the QDPM extends the Hilbert space of the QDM to include both fully-packed hard-core dimer configurations, as well as those with pentamers--four dimers touching a vertex. We define a local Hamiltonian that provides the simplest non-trivial dynamics that preserve this local constraint. Consequently, the Hamiltonian has an exact local $Z_3$ symmetry and conserves a $Z_3$ topological winding number; thus the QDPM has a 9-fold topological ground state degeneracy on the torus.

Additionally, we present a numerical study of the ground state of the QDPM at an exactly soluble point. Using Monte Carlo calculations, we explicitly demonstrate the absence of symmetry breaking order and the exponential decay of correlation functions. Additionally, we define a $Z_3$ magnetic string operator, whose correlations display no evidence of magnetic vortex condensation. Finally, by a studying systems doped with a pair of  monomer defects, we demonstrate that monomers are deconfined in the QDPM. Thus we argue that the ground state of the QDPM is a $Z_3$ topologically ordered dimer liquid.

This paper is organized as follows. In \secref{sec:Background} we provide relevant background for the QDM and related models. In \secref{sec:QDPM} we present the details of the QDPM. \secref{sec:numerics} provides the results of a Monte Carlo study of the ground state of the QDPM at its exactly soluble point. Finally, \secref{sec:discussion} discusses potential future studies of the QDPM.

\section{Background}
\label{sec:Background}

\subsection{Generalized quantum dimer models}
\label{sec:GQDM}

\begin{figure}[htpb]
    \centering
    \includegraphics[width=0.9\linewidth]{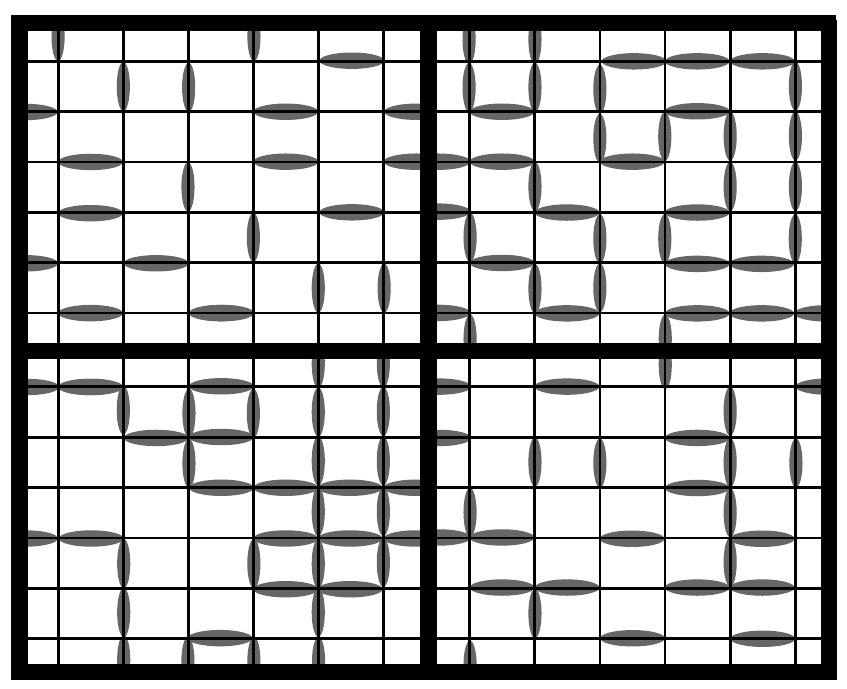}
    \caption{Examples of GQDMs with different local constraints. Clockwise from the upper left panel; fixed dimer number: $n_d=1$, $n_d=2$;  fixed dimer parity: odd, even.}
    \label{fig:example_local_constraints}
\end{figure}
Here we will provide background about generalized quantum dimer models on the square lattice. We consider systems with Ising-like dimer degrees of freedom on each link, such that each link is either unoccupied or occupied by a single dimer. Given an Ising spin $\sigma_\ell$ on each link $\ell$ we define the number of dimers on the link $n_\ell \equiv 1/2+\sigma_\ell^z$, where $\sigma_\ell^i$ are the associated Pauli matrices. The Hilbert space is spanned by an orthonormal basis of dimerizations of the square lattice that satisfy a local constraint at each vertex. We define a $U(1)$ charge $Q_v$ at each vertex $v$ of the lattice in terms of the number of dimers touching $v$, $n_v\equiv \sum_{\ell \in v}n_\ell$:
\begin{equation}
Q_v \equiv \pm \left( n_v - c_P \right) \notag
\end{equation}
where $c_P = 2$, ($c_P=1$) for models in the even (odd) sector of the constraint, and the $+$($-$) sign corresponds to vertices on the $A$($B$) sublattice. A local dimer number constraint requires $Q_v$ to vanish for all vertices; a dimer parity constraint requires $(Q_v\mod 2)$ to vanish.
 
The Hamiltonian of a GQDM is of the form
\begin{equation}
H_{\rm{GQDM}} = -\sum_\nu t_\nu  {T}_{\nu} + \sum_\nu v_\nu  {V}_{\nu} \label{eq:HGQDM}
\end{equation}
where ${T}_\nu$ are off-diagonal kinetic energy operators, ${V}_\nu$ are diagonal potential energy operators, $t_\nu$ and $v_\nu$ parametrize the strength of these terms, and each $\nu$ represents a set of  neighboring links. We will limit our discussion to $t_\nu >0$. ${T}_\nu$ is constructed to generate resonances between two distinct allowed dimerazations of $\nu$, $c_\nu$ and its compliment $\bar{c}_\nu$, and is written as:
\begin{equation}
 {T}_{\nu} = \eket{c_\nu}\ebra{\bar{c}_\nu} + h.c.;  \notag
\end{equation}
in the simplest case, $ {T_\nu}$ involves the minimum non-trivial dimer dynamics that preserve the constraint. Similarly, the potential energy term provides an energy cost for states with these local configurations:
\begin{equation}
 {V}_{\nu} = \eket{c_\nu}\ebra{c_\nu} +  \eket{\bar{c}_\nu}\ebra{\bar{c}_\nu}.  \notag
\end{equation}
Therefore we may define a Hamiltonian for a GQDM using \Eqref{eq:HGQDM} and defining a set of dimer resonances $\{(c_\nu,\bar{c}_\nu)\}$ and their corresponding strengths $\{(t_\nu,v_\nu)\}$:
\begin{equation}
\HGQDM \biggl( \bigl\{\left(c_\nu,\bar{c}_\nu\right)\},\{\left(t_\nu,v_\nu\right)\bigr\}\biggr).  \notag
\end{equation}
 In general, $[T_\nu, T_{\nu'}] \neq 0 $ and $[T_\nu,V_{\nu'}]\neq0$ when $\nu$ and $\nu'$ share links, and thus the ground state of $H_{\rm{GQDM}}$ is not explicitly known.

However, at the special points in the $\{(t_\nu,v_\nu)\}$ parameter space, known as the Rokhsar-Kivelson (RK) points, when \Eqref{eq:HGQDM} can be written as the sum of projection operators
\begin{equation}
H_{\rm{RK}} = \sum_\nu  {h}_\nu \label{eq:HRK},
\end{equation}
where all $h_\nu$ are projectors with eigenvalues $\{0,1\}$, the exact ground state is known to be
the zero energy state of \Eqref{eq:HRK} that is simultaneously annihilated by all ${h}_\nu$. Such a zero energy state, known as the RK state, may be formed from an equal superposition of all dimerizations that are connected by $\{T_\nu\}$:
\begin{equation}
\eket{\Phi_{\RK}^{\Omega}} = \sum_{C \in \Omega} \eket{C}  \notag
\end{equation} 
where $\Omega$ is a subset of the Hilbert space connected by $\{T_\nu\}$. Thus $\{ \ket{\Phi_{\RK}^\Omega} \}$ are the exact, degenerate ground states of $H_{\rm{GQDM}}$ with a degeneracy labelled by $\Omega$ at the RK point.

Because of the local constraints, all states $\ket{\psi}$ in the Hilbert space of a GQDM are invariant under local gauge transformations $G_v \ket{\psi} = \ket{\psi}$, $G_v$ of the general form~\cite{Moessner2001}
\begin{equation}
G_v = e^{i \alpha_v Q_v}, \label{eq:Gv}
\end{equation}
where the allowed values of $\alpha_v$ depend on the constraint.  
By construction, the Hamiltonian is gauge invariant and thus commutes with $G_v$
\begin{equation}
\left[ H_{\rm{GQDM}}, G_v \right] = 0. \notag
\end{equation}

If a defect is introduced which violates the local constraint at a vertex, this can be viewed as an ``electric charge'' living on the vertex. One may determine if such electric charges are confined by introducing a pair of such defects and computing their spatial correlation function; if the free energy cost of defects at large distances remains finite, electric charges are deconfined in the GQDM ground state.

\begin{figure}[tpb]
    \centering
    \includegraphics[width=1.0\columnwidth]{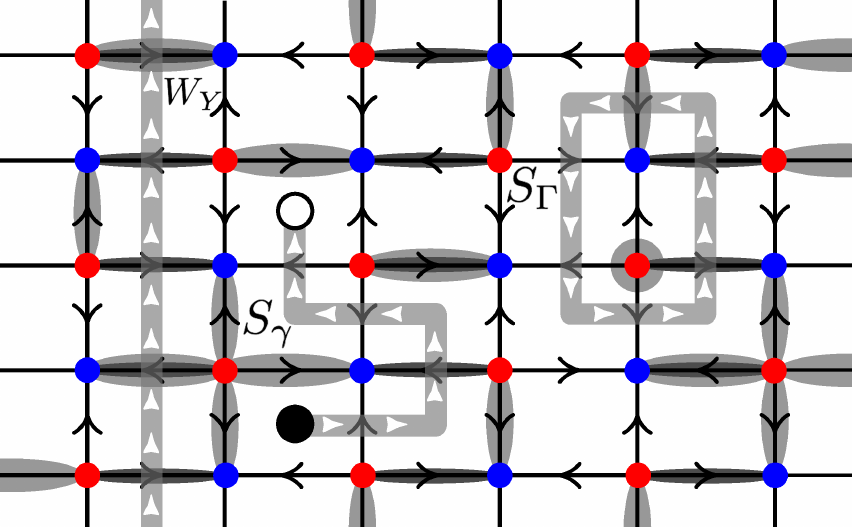}
    \caption{The A(B) sublattices of the square lattice are colored red (blue) and the link orientations are chosen to point from the A to the B sublattice. The narrow dark dimers are a static background dimerization used to fix the gauge, in this case chosen to be a columnar pattern; the wider light gray dimers are the physical degrees of freedom. $W_Y$ is a magnetic string operator with a non-trivial winding around one axis of the torus that measured the topological sector; $S_\gamma$ and $S_\Gamma$ are examples of open and closed string operators, respectively. Note the monomer defect which can be detected by $S_\Gamma$.}
    \label{fig:orientedSL}
\end{figure}

The gauge symmetry allows for the construction of a gauge invariant magnetic string operator $S_\gamma$ along an oriented path $\gamma$ path on the dual lattice. We first orient all the links of the square lattice such that they point from the $A$ sublattice to the $B$ sublattice (see \figref{fig:orientedSL}) and define an oriented dimer flux $\Phi_\gamma$ through the path:
\begin{equation}
\Phi_\gamma \equiv \sum_{\ell \in \gamma} \rm{sgn}\left(\ell,\gamma\right) n_\ell \label{eq:phi}
\end{equation}
where $\mathrm{sgn}(\ell,\gamma)$ is positive (negative) if $\ell$ is oriented to the right (left) with respect to the orientation of $\gamma$. Additionally, we define background flux $\tilde{\Phi}_\ell$
\begin{equation}
\tilde{\Phi}_\gamma \equiv  \sum_{\ell \in \gamma} \rm{sgn}\left(\ell,\gamma\right) \tilde{n}_\ell \label{eq:phiBackground}
\end{equation}
where $\tilde{n}_\ell$ is the number operator for a static background dimer configuration which is chosen to fix the gauge. With this gauge fixing, the total charge enclosed in a closed loop $\Gamma$ is $Q_\Gamma = \Phi_\Gamma- \tilde{\Phi}_\Gamma$. An alternative physical picture is that there are static charges of charge $+c_P$ ($-c_P$) sitting at the vertices of the $A$($B$) sub-lattice, which are the source of the non-vanishing $\Phi_\Gamma$ flux through a closed loop.

We now may define the string operator
\begin{equation}
S_\gamma \left( \alpha \right) \equiv e^{i \alpha \left( \Phi_\gamma - \tilde{\Phi}_\gamma \right)} \label{eq:Sgamma}
\end{equation}
where $\alpha$ such that $S_\gamma$ commutes with $\HGQDM$ everywhere except at the ends of $\gamma$. Therefore, for closed loops $\Gamma$, $[ S_\Gamma,\HGQDM]=0$. For open strings, $S_\gamma$ does not commute with the Hamiltonian and can be viewed as creating a pair of ``magnetic charges'' at the ends of the string: correspondingly $S_\gamma \ket{\Phi_{\RK}}$ can used as a trial wavefunction for an excited state with a pair of magnetic vortices. If a state has topological order with deconfined electric charges, these magnetic vortices should be gapped excitations, and correspondingly the magnetic string operator should decay exponentially~\cite{Read1989a,Senthil2000,Senthil2001e}. The condensation of magnetic vortices can drive a phase transition from a topologically ordered phase to a conventionally order phase~\cite{Jalabert1991,Ralko2007,Huh2011}. Thus, the magnetic string correlation can be used as test for deconfinement of electric charges.

On surfaces with non-trivial topology, we may define the topological loop operators
\begin{equation}
W_{X} \left( \alpha \right) \equiv S_{\Gamma_{X}} \left( \alpha \right)  \notag
\end{equation}
where $\Gamma_{X}$ is a topologically non-trivial closed loop that winds once about the $x$ axis; we define $W_Y$ similarly. Since $W_{X} (\alpha)$ and $W_{Y} (\alpha)$ commute with the Hamiltonian, we may use their eigenvalues to label the ground states.

\subsection{Fixed dimer number constraints}

The two unique fixed-number constraints on the square lattice lead to the QDM in the odd sector ($n_v = 1$) and the quantum full-packed loop model (QFPLM) in the even sector ($n_v = 2$). For configurations that satisfy these number constraints, $Q_v$ strictly vanishes, and thus all states in the Hilbert space are invariant under $G_v$ for $0 \leq \alpha_v \leq 2 \pi$ in \Eqref{eq:Gv}; this represents a local $U(1)$ gauge symmetry~\cite{Moessner2001}. 

The Hamiltonian for these models gives dynamics to parallel dimers on the plaquettes of the lattice:
\begin{equation}
\HQDM\left(t,v\right) \equiv \HGQDM \biggl( \bigl\{ \left(\vrtplqt,\hrzplqt\right) \bigr\},\bigl\{ \left(t,v\right) \bigr\} \biggr)  \label{eq:HQDM}
\end{equation}
where we have chosen $t_\bplqt = t$, $v_\bplqt =v, \forall \bplqt$. In this case of a fixed number constraint, the Hamiltonian conserves $Q_v$ and therefore the dimer flux through any closed loop such that:
\begin{equation}
\left[ \Phi_\Gamma, \HQDM \right] = 0.  \notag
\end{equation}
We can define two conserved topological winding numbers $\Phi_{X}$ \& $\Phi_{Y}$ which are the dimer flux through topologically non-trivial loops that wind around each axis of the lattice when the system has periodic boundary conditions (see \figref{fig:orientedSL}). On a lattice of dimensions $L_x\times L_y$, the eigenvalues of $\Phi_X$ and $\Phi_Y$ are integers in the range
\begin{equation}
-L_{x}/2 \leq  \Phi_{X} \leq L_{x}/2,  \notag
\end{equation}
and similarly for $\Phi_Y$, thus there are $\mathcal{O} (L_x L_y)$ distinct topological sectors. 

At the Rokhsar-Kivelson point $t=v$, we may write $\HQDM$ in the form of \Eqref{eq:HRK} by defining projection operators  $ {h}_\nu $:
\begin{equation}
 {h}_\nu \equiv \biggl( \eket{ c_\nu} - \eket{\bar{c}_\nu} \biggr)\biggl( \ebra{ c_\nu} - \ebra{\bar{c}_\nu} \biggr).  \notag
\end{equation} 
Since the plaquette flip term in \Eqref{eq:HQDM} is believed to be ergodic in each of topological sector $(\Phi_x,\Phi_y)$ on the square lattice, each topological sector corresponds to a unique RK ground state where $\Omega$ is labeled by $(\Phi_x,\Phi_y)$. Thus the  RK points of the QDM and QFPLM have an \emph{extensively} degenerate ground states.

Defects in these models violate the local dimer number constraint and thus carry non-zero U(1) electric charge. \tabref{tab:defects} lists all the possible defects and their associated U(1) charges in QDM and QFPLM. When a defect lives on sublattice A, it has the opposite charge that the same defect on sublattice B has. Note that in the QDM, the maximum charge is $\pm 3$.

The QDM and QFPLM on the square lattice have isolated liquid ground states at the RK point surrounded by symmetry broken phases~\cite{Rokhsar1988,Leung1996,Syljuasen2005,Syljuasen2006,Syljuasen2004,Moessner2008}. The RK point of these models have been shown to have power-law decaying dimer-dimer and monomer-monomer correlation functions and gapless excitations~\cite{Fisher1961,Kasteleyn1961,Fisher1966a,Fisher1963,Kasteleyn1963a,Sutherland1988b,Sutherland1988a,Kohmoto1988a,Liang1988,Sutherland1988,Kohmoto1988,Levitov1990,Leung1996,Henley1997,Henley1997,Henley2004a,Pollmann2006}. The algebraic decay of monomer correlations is indicative of the logarithmic confinement of electric charges in these models. Thus, such fixed number constrained models on bipartite lattices are expected to have an isolated unstable gapless critical quantum liquid points with logarithmically confined defect and an extensive topological degeneracy. These features of fixed number constrained models are consistent with the effective U(1) gauge theory description of the RK point~\cite{Fradkin2004,Fradkin1990,Moessner2001}; in 2+1D pure U(1) gauge theories are confining and thus the lack of an extended deconfined topologically ordered phase near the RK point~\cite{Polyakov1987}.

\begin{table}
  \begin{tabular}{| l | c | c | c | c | c |}
  	\hline
    type & $n_v$& configuration & $Q_v$ & $Q_v^{Z_2}$  & $Q_v^{\rm{Z_3}}$  \\
	\hline
    monomer & $0$& $\svrtmon$ & -1 & 1 &  -1   \\
	\hline
    dimer  & $1$ & $\svrtdmr$& 0 & 0 &   0  \\
	\hline
    trimer  & $2$ & $\svrttrmr$ $\svrttrmrtwo$& 1 & 1 &   1  \\
	\hline
    tetramer & $3$& $\svrtquat$  & 2 & 0 &   -1  \\
	\hline
    pentamer & $4$& $\svrtpent$ & 3 & 1 &  0   \\
	\hline
  \end{tabular}
  \caption{Local dimer configurations where $n_v$ dimers touch a vertex; only one representative of symmetry related configurations is shown. $Q_v$, $Q_v^{Z_2}$, and the $Q_v^{Z_3}$ are the associated $U(1)$, $Z_2$ and $Z_3$ charges on the A sub-lattice in the odd sector theories (QDM, odd IGT, QDPM). Defects on the B sub-lattice have opposite charge, with the exception of $Q_v^{Z_2}$, as there is only one type of non-trivial $Z_2$ charge.}
  \label{tab:defects}
\end{table}

\subsection{Fixed dimer parity constraints}

We now relax the local constraint to a parity constraint such that only $Q_v^{Z_2} \equiv (Q_v \mod 2)$ is required to vanish; the resulting GQDM are the even and odd sectors are the  even \& odd Ising gauge theories (IGT) with constraints $n_v \in \{0,2,4\}$ and $n_v \in \{1,3\}$, respectively. In these cases, the local constraint requires $\alpha_v \in \{ 0, \pi\}$ in \Eqref{eq:Gv} and therefore the gauge symmetry is reduced from $U(1)$ to $Z_2$. The set of allowed dimer resonances $\mathcal{C}_{\rm{IGT}}$ is expanded accordingly:
\begin{align}
\mathcal{C}_{\rm{IGT}} &\equiv \bigl\{ \left(c_{\Box,i},\bar{c}_{\Box,i}\right) \bigr\} \\
&=\bigl\{
\left(\vrtplqt,\hrzplqt\right),\left(\threeleftplqt,\oneleftplqt\right),\left(\twodownplqt,\twoupplqt\right),\left(\bplqt,\fplqt\right) \bigr\}
\end{align}
where we implicitly include symmetry related resonances. We will limit our discussion to $v_\bplqt=0$ such that the corresponding Hamiltonian is
\begin{equation}
\HIGT\left(t\right) \equiv \HGQDM \biggl( \mathcal{C}_{\rm{IGT}},\bigl\{ \left(t,0\right) \bigr\} \biggr).  \notag
\end{equation}
In this case the equal superposition RK wavefunction is the exact ground state and occurs in the absence of a potential energy term, as we can define a projection operator on each plaquette
\begin{equation}
h_\Box \equiv \frac{1}{2} \left( \mathbb{1} - \sum_i \eket{c_{\Box,i}}\ebra{\bar{c}_{\Box,i}} + h.c. \right)  \notag
\end{equation}
such that $\HIGT$ takes the form of \Eqref{eq:HRK} (up to a constant), and the RK state is the ground state for all values of $t$.

The additional dynamics of $\HIGT$ vs. $\HQDM$ no longer conserve the dimer flux $\Phi_\Gamma$ through a closed curve; only $\Phi_\Gamma \mod 2$ is preserved by $\HIGT$. Consequently, to define the string and loop operators that commute with $\HIGT$ (for closed strings) we fix $\alpha = \pi$ in \Eqref{eq:Sgamma}:
\begin{equation}
S_\gamma^{Z_2} \equiv S_\gamma \left( \pi \right), \quad W_{X}^{Z_2} \equiv W_{X} \left( \pi \right), \notag
\end{equation}
and $W_Y^{Z_2}$ is defined similarly. Thus the $W_{X}^{Z_2}$ and $W_{Y}^{Z_2}$ have two eigenvalues $\pm 1$, and the topological sectors on the torus can be labeled as~$\Omega= (\pm 1,\pm1)$; consequently the extensive degeneracy of the fixed number constrained models has been reduced to a finite ground state degeneracy in the fixed parity models.

Defects violating the dimer parity constraint in the IGT's have $Q_v^{Z_2} = 1$, as listed in \tabref{tab:defects}; we identify these defects as $Z_2$ electric charges on the corresponding vertices. Note that there is only one non-trivial charge in $Z_2$ gauge theory, thus there is no distinction between the charges of defects living on the two sublattices.

The ground states of $\HIGT$ are fourfold degenerate on the torus, and have exponentially decaying dimer-dimer and monomer-monomer correlations and a finite energy gap; such features demonstrate the deconfined topologically ordered nature of the ground state of the IGTs.

We may understand the emergence of the $Z_2$ topologically ordered phase of IGTs from the QDM and QFPLM from the following picture. The additional vertex configurations that are allowed in the IGTs relative to the models with a dimer number constraint  have U(1) charge $Q_v = \pm 2$ (see \tabref{tab:defects}). It is known that when a $U(1)$ gauge field is coupled to a dynamic charge $q$ matter field, this results in a $Z_q$ gauge theory when the matter field condenses~\cite{Fradkin1979}; as in the IGT's we have added dynamic charge 2 defects to the underlying $U(1)$ symmetric model, and we may expect a $Z_2$ gauge theory to result. We note that pentamers are charge $\pm3$ objects in the QDM (see \tabref{tab:defects}); thus below we investigate whether introducing pentamers into the QDM leads to an effective $Z_3$ gauge theory (see \appref{sec:Z3GT} for a review of $Z_3$ gauge theory).

\section{The Quantum Dimer Pentamer Model}
\label{sec:QDPM}

We define the Hilbert space of the QDPM to be a GQDM where vertices are constrained to be touched by a single dimer or the center of a pentamer, such that $n_v \in \{1,4\}$. For defect-free configurations in the QDPM, $Q_v$ only vanishes modulo $3$;  thus we may define a $Z_3$ charge,
\begin{equation}
Q_v^{Z_3} \equiv \left(Q_v~\rm{mod}~3\right)  \notag
\end{equation}
 which strictly vanishes at each vertex, representing the local constraint. All states in the QDPM are invariant under an exact local $Z_3$ gauge symmetry given by \Eqref{eq:Gv} with $\alpha_v \in \{0,\pm 2\pi/3\}$. Thus introducing pentamers in the QDM reduces the gauge symmetry from $U(1)$ to $Z_3$.

The Hamiltonian of the QDPM is defined by \Eqref{eq:HGQDM} including both plaquette flips as well as local dimer resonances that create and destroy pairs of pentamers, and allow single pentamers to translate on the same sub-lattice; these local dynamics involving pentamers are illustrated in  \tabref{tab:pentamer_moves}. Note that pentamers give dynamics to ``staggered" regions with no flippable plaquettes; thus we expect the QDPM to have weaker correlations that the QDM, where the dynamics are more constrained. We may write $\HQDPM$ as the following:
\begin{equation}
\HQDPM \equiv \HQDM + H_{\rm{pen}}  \notag
\end{equation}
where $H_{\rm{pen}}$ only involves the pentamer resonances in \tabref{tab:pentamer_moves} and those related by symmetry:
\begin{align}
H_{\rm{pen}} \equiv \HGQDM \biggl( &\bigl\{ \left(c_{\nu_1},\bar{c}_{\nu_1}\right),\left(c_{\nu_2},\bar{c}_{\nu_2}\right),\left(c_{\nu_3},\bar{c}_{\nu_3}\right) \bigr\},\biggr.\notag\\
\biggl. &\bigl\{ \left(t_1,v_1\right),\left(t_2,v_2\right),\left(t_3,v_3\right) \bigr\} \biggr)  \label{eq:Hpen}
\end{align} 
 
\begin{table}
    \begin{tabular}{| l | c | c |}
  	\hline
    $\nu$ & $c_\nu$ & $\bar{c}_\nu$ \\
	\hline
    $\nu_1$ & $\pentmvcreate$ &  $\pentmvanal$  \\
	\hline
    $\nu_2$ &$\penmvhrzleft$&$\penmvhrzright$ \\
	\hline
    $\nu_3$ &$\penmvdiagup$& $\penmvdiagdown$\\
	\hline
  \end{tabular}
  \caption{The pairs of local dimer \& pentamer configurations $c_\nu\leftrightarrow\bar{c}_\nu$ generate minimal local dynamics that create/annihilate pairs ($\nu_1$) of pentamers and translate single pentamers ($\nu_2$ \& $\nu_3$). Note that pentamers always stay on the same sublattice which is associated with their $U(1)$ charge. }
  \label{tab:pentamer_moves}
\end{table}

Because of the terms involving pentamers, $\HQDPM$ conserves $Q_v^{Z_3}$ rather than $Q_v$, and thus $\Phi_\Gamma$ does not commute with the Hamiltonian for closed loops $\Gamma$. The appropriate string and loop operator corresponds to \Eqref{eq:Sgamma} with $\alpha = 2\pi/3$:
\begin{equation}
S_\gamma^{Z_3} \equiv S_\gamma \left( 2\pi/3\right), \quad W_{X}^{Z_3} \equiv W_{X} \left( 2\pi/3\right) \label{eq:SQDPM}
\end{equation}
and similarly for $W_Y^{Z_3}$. With these definitions, closed string operators $S_\Gamma^{Z_3}$ and the topological loop operators $W_{X}^{Z_3}$ and $W_{Y}^{Z_3}$ commute with $\HQDPM$.

Our numerical simulations (discussed below in \secref{sec:numerics}) suggest that the dynamics of $\HQDPM$ are ergodic in each topological sector $\Omega$ that are labeled by the topological loop operators:
\begin{equation}
\Omega = \left(W_X^{Z_3},W_Y^{Z_3} \right).  \notag
\end{equation}
The exact zero energy ground state at the RK point, where $t_i=v_i, i \in \{ 1,2,3\}$, is the RK state, where the superposition is taken to be over all states in the topological sector. Since the $W_{X}^{Z_3}$ and  $W_{Y}^{Z_3}$ have eigenvalues $\{1,e^{i2\pi/3},e^{-i2\pi/3}\}$, the RK point of the QDPM is therefore 9-fold degenerate on the torus.

Defects in the QDPM include monomers, trimers, and tetramers. As these dimer configurations violate the local constraint at the vertex, these defects carry a non-vanishing $Z_3$ electric charge $Q_v^{Z_3} = \pm 1$, where the sign depends on the sub-lattice and type of defect. \tabref{tab:defects} lists the $Z_3$ charge for these defects on one sublattice; defects of the same type carry opposite charge when living on opposite sublattices.

We have demonstrated that the ground state at the RK point of the QDPM has an exact $Z_3$ gauge symmetry and a corresponding $9$-fold topological degeneracy on the torus. We emphasize that despite the exact $Z_3$ local gauge symmetry of the QDPM, there is not an exact mapping to a $Z_3$ gauge theory on the square lattice, as the QDPM has Ising degrees of freedom on the links, rather than the 3 dimensional local Hilbert space of a $Z_3$ gauge theory (see \appref{sec:Z3GT} for details of $Z_3$ gauge theory).  It remains to address whether this RK state is in the deconfined $Z_3$ topologically ordered phase of a $Z_3$ gauge theory, or a confined symmetry broken or critical state, which we discuss below.

\section{Numerical Study of the RK ground state of the QDPM}
\label{sec:numerics}

To determine the nature of the ground state at the RK point of the QDPM, we have Monte Carlo sampled $\ket{\Psi_{\rm{RK}}}$, using local updates corresponding to the dimer resonances shown in \tabref{tab:pentamer_moves} and single plaquette flips, as defined in \Eqref{eq:HQDM}. Below we present the results of calculations on $L\times L$ square lattices with periodic boundary conditions on systems of linear dimension up to $L=128$.

\subsection{Topological sectors}

\begin{figure}[]
    \centering
    \includegraphics[width=1.0\linewidth]{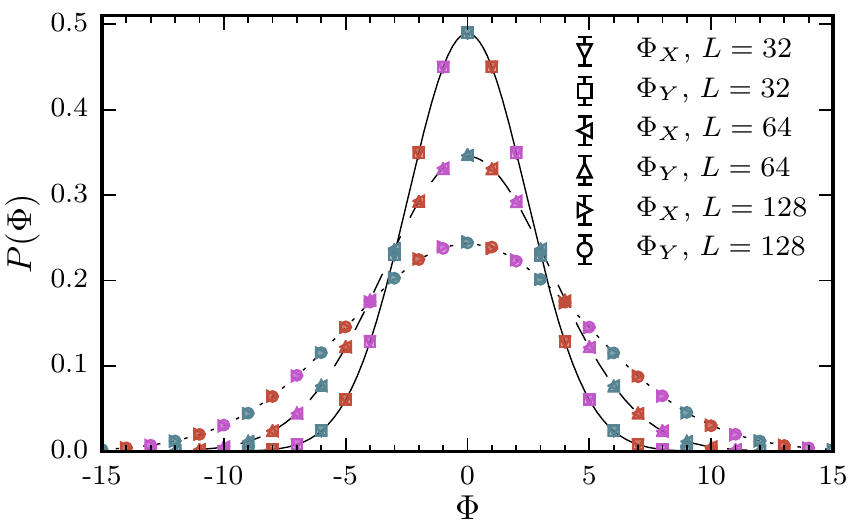}
    \caption{The histogram of the $U(1)$ topological winding numbers $\Phi_X$ and $\Phi_Y$ as measured in the ground state of the RK point of the QDPM for three different lattice sizes. The colors correspond to calculations initialized to different topological sectors $(\Phi_X,\Phi_Y)$: blue $(0,0)$, red $(1,1)$, and purple $(-1,-1)$. The lines are fits to a Gaussian for each system size. Note that $\Phi$ is conserved modulo $3$, resulting in a $9$-fold topological degeneracy on a torus.}
    \label{fig:u1_wind_qdpm}
\end{figure}

We explicitly demonstrate that the extensive topological degeneracy of the QDM is reduced to a finite degeneracy in the QDPM by computing the histogram of $U(1)$ topological winding numbers $\Phi_{X}$ and $\Phi_{Y}$, which is shown in \figref{fig:u1_wind_qdpm}. The different colors represent calculations that were initialized to different winding sectors $(\Phi_X,\Phi_Y)$, and we find that $\Phi_X$ and $\Phi_Y$ are not  strictly conserved (as they are in the QDM), but are only conserved modulo $3$. The histogram $P(\Phi)$ has a Gaussian shape with a width that increases with system size; this is expected as $\Phi = \pm L/2$ correspond to ordered configurations and thus there are fewer dimer configurations near these limits. We find no evidence of a lack of ergodicity within each of the 9 topological sectors. For the remainder of the paper we present results in the $(0,0)$ topological sector, but we see no qualitative dependence between topological sectors for any of the results presented here.  Thus we find that the QDPM has a \emph{finite} $9$-fold degeneracy on the torus, where the topological sectors are labeled $(W_X^{Z_3},W_Y^{Z_3})$, in contrast with the extensive degeneracy of the QDM.

\subsection{Dimer density correlations}

\begin{figure}[t]
        \centering
        \includegraphics[width=1.0\columnwidth]{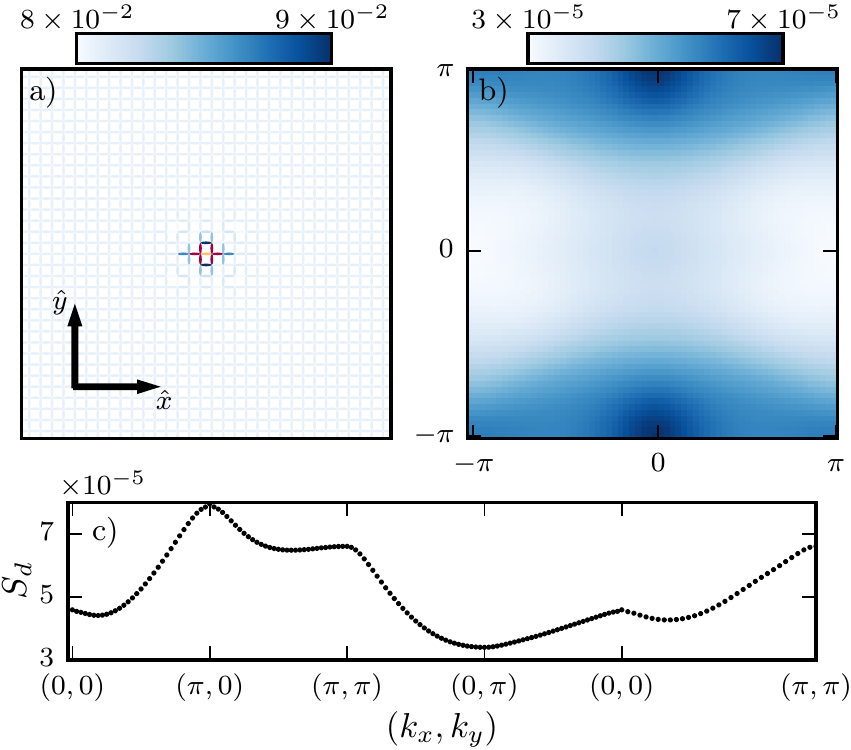}
        \caption
        {
            (a) Gray scale of the dimer-dimer correlation function $\langle n_{\bm{0}}
            n_{\bm{r}}\rangle$ on an $L=32$ lattice. The origin corresponds to the orange link, and
            $\bm{r}$ varies over all other links of the lattice. Red links have correlations that are off
            the given scale. 
            (b) The horizontal dimer density structure factor $S_d(\bm{k})$ for the
            QDPM RK ground state on an $L=64$ lattice. The lack of $x\leftrightarrow y$ rotational
            symmetry is due to the fact that horizontal links have been isolated. 
            (c) The horizontal dimer-dimer structure factor $S_d(\bm{k})$ shown along a
            path in $k$ space for an $L=64$ lattice. Statistical errors are smaller than the symbols.
        }
        \label{fig:dimer_3way}
    \end{figure}

\begin{figure}[t]
    \centering
    \includegraphics[width=1.0\columnwidth]{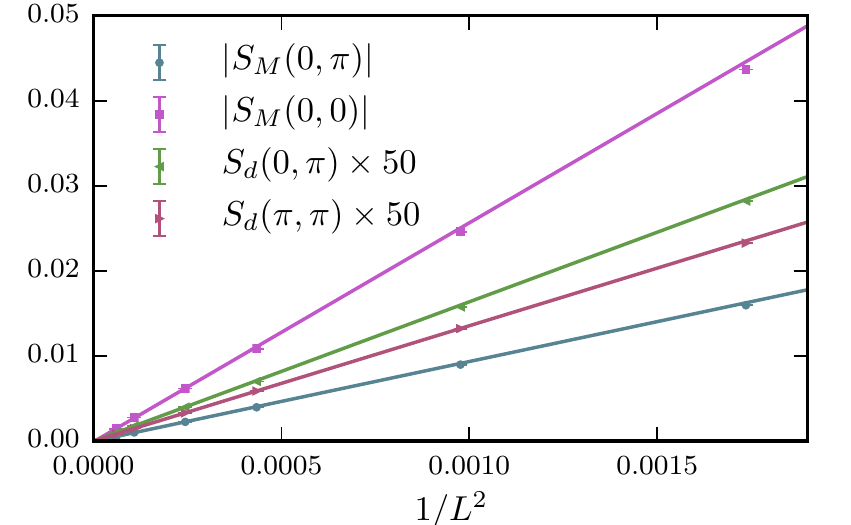}
    \caption{Finite-size scaling of the maxima in the dimer ($S_d$) and magnetic string ($S_M$) structure factors. The lines are fits to $L^{-2}$ scaling. All maxima vanish in the thermodynamic limit, indicating that the RK state of the QDPM is a symmetric dimer liquid. }
    \label{fig:bragg_scaling}
\end{figure}

We will now present a numerical study of the dimer density correlations in the RK ground state of the QDPM. In \figref{fig:dimer_3way} (a) we plot the bare dimer-dimer density correlation function $\langle n_{\bm{0}} n_{\bm{r}}\rangle$ on a $L=32$ lattice; here the origin $\bm{0}$ is represented by the orange link, and $\bm{r}$ varies over all other links of the lattice. These bare dimer correlations display only short range correlations with no obvious symmetry breaking. In \figref{fig:dimer_3way} (b) we show the equal-time dimer-dimer structure factor, 
\begin{equation}
S_d\left( \bm{k} \right) \equiv \frac{1}{ L^4} \sum_{i,j} e^{-i \bm{k} \cdot \left(\bm{r}_i - \bm{r}_j\right)} \left \langle n_{\bm{r}_i} n_{\bm{r}_j} \right \rangle  \notag
\end{equation}
where we restrict the sum to horizontal links, for a $L=64$ lattice. 
\figref{fig:dimer_3way} (c) shows $S_d(\bm{k})$ for a particular path through the Brillouin zone. The dimer structure factor displays broad maxima at $\bm{k} = (0, \pm \pi)$ and $\bm{k} = (\pm \pi, \pm \pi)$. The lack of rotational symmetry of $S_d(\bm{k})$ is due to the fact that we have isolated horizontal dimers; the correlations between vertical dimers display the same behavior, rotated by $\pi/2$.
We have plotted the finite size scaling of the intensity of the broad maximum of $S_d$ in  \figref{fig:bragg_scaling}, which we find to vanish in the thermodynamic limit; these maxima are therefore consistent with finite size effects due to short distance correlations. In \appref{sec:dimer_order_parameters} we additionally have plotted order parameters that specifically test for rotational and translational symmetry breaking; in all cases we find no symmetry breaking in the thermodynamic limit. 


\begin{figure}
    \centering
    \includegraphics[width=1.0\columnwidth]{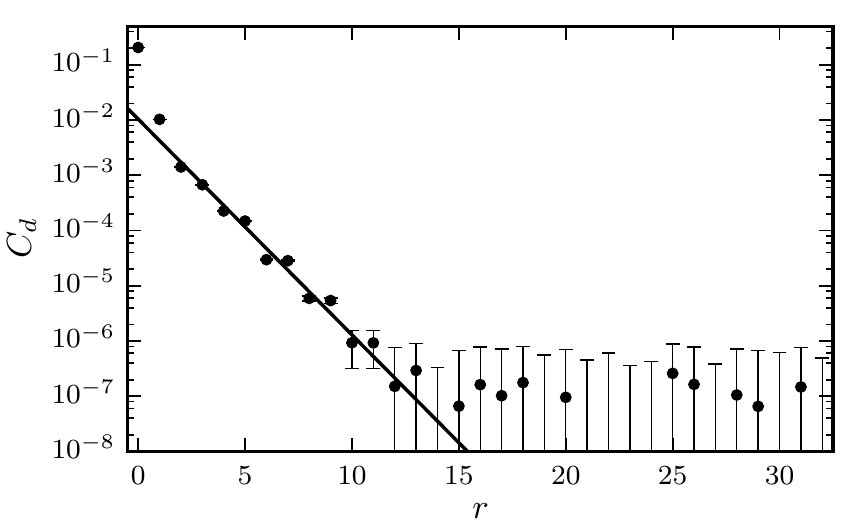}
    \caption{The dimer-dimer density correlation function $C_d$ between parallel links in the QDPM RK ground state on a $L=64$ lattice, along the direction perpendicular to the link orientation.  The lines represent the best fit to an exponential decay with correlation length $\chi\approx 1$.}
    \label{fig:spatial_dmr_cor}
\end{figure}

Additionally we have studied the nature of the decay of the connected dimer-dimer correlation function,
  \begin{equation}
    C_{\mathrm{d}} \left(r\right) \equiv \langle n_{\bm{0}} n_{\bm{r}} \rangle - \langle n_{\bm{0}} \rangle   \langle n_{\bm{r}} \rangle   \notag
\end{equation}
where $r$ is the distance between the links $\bm{0}$ and $\bm{r}$. In \figref{fig:spatial_dmr_cor}, $C_d$ between parallel links is plotted as a function of separation, for $\bm{r}$ perpendicular to the orientation of the dimers for a $L=64$ lattice. We find that $C_d$ decays exponentially with distance having a correlation length  $\chi\approx 1$. Additionally, $C_d$ displays a sub-lattice oscillation that is consistent with the finite-size maximum in $S_d$ at $(0,\pi)$. Thus we find that the RK point of the QDPM is a symmetric dimer liquid with exponentially decaying dimer correlations, in contrast with the power-law decaying correlations of the QDM.

\subsection{$Z_3$ magnetic string correlations}

\begin{figure}
        \centering
        \includegraphics[width=1.0\columnwidth]{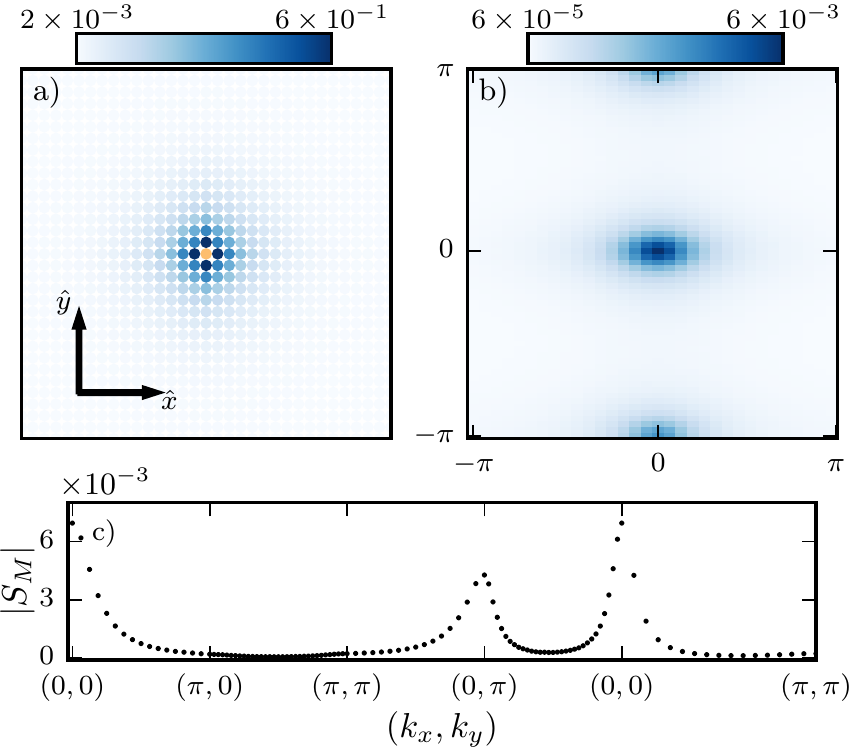}
        \caption
        {
            (a) Color scale of the $Z_3$ magnetic string correlation function $|C_M(\bm{r}_{\bm{0}p})|$,
            on a $L=32$ lattice where $\bm{0}$ corresponds to plaquette represented by the orange circle.
            (b) The magnitude of the $Z_3$ magnetic string structure factor $S_M$ for a $L=64$ lattice, using a columnar background dimerization as shown in \figref{fig:orientedSL}.             
            (c) $|S_M|$ shown along a path in $k$ space. Statistical errors are smaller than the symbols.
        }
        \label{fig:vison_3way}
    \end{figure}

To investigate the $Z_3$ gauge theory description of the QDPM RK ground state, we study the $Z_3$ magnetic string correlation function as defined in \eqref{eq:Sgamma} \& \eqref{eq:SQDPM}:
\begin{equation}
C_M \left( \bm{r}_{pp'} \right) \equiv \left \langle \exp \left[ \frac{2 \pi}{3} i \left( \Phi_{\gamma_{pp'}} -\tilde{\Phi}_{\gamma_{pp'}} \right) \right] \right \rangle  \notag
\end{equation}
where $p$ and $p'$ are two plaquettes, $\bm{r}_{pp'}$ is the displacement between $p$ and $p'$,  and $\gamma_{pp'}$ is any open string on the dual lattice
connecting $p$ and $p'$. In \figref{fig:vison_3way} (a), we have plotted $|C_M|$ for a $L=32$ lattice. Note that while $C_M$ depends on the choice of the background dimerization which fixes the gauge, $|C_M|$ is independent of the background dimerization. We find only short range magnetic string correlations that display no obvious symmetry breaking. 

\begin{figure}
    \centering
    \includegraphics[width=1.0\columnwidth]{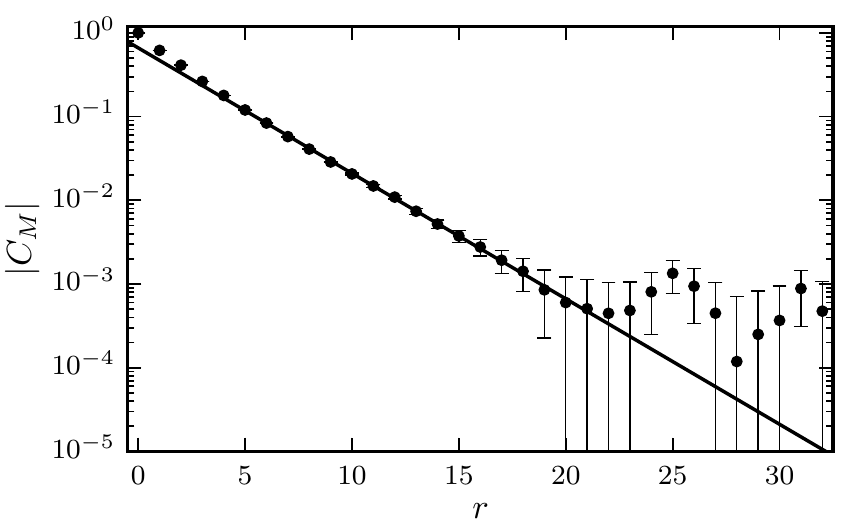}
    \caption{ The magnitude of the $Z_3$ magnetic string correlation $|C_M|$ along one axis of the dual lattice of a $L=64$ lattice. The line corresponds to an exponential fit with correlations length $\chi\approx 3$.}
    \label{fig:vison_cor}
\end{figure}

Additionally, we have computed the magnetic string structure factor
\begin{equation}
S_M\left( \bm{k} \right) \equiv \frac{1}{ L^4} \sum_{p,p'} e^{-i \bm{k} \cdot \bm{r}_{pp'}} C_M\left( \bm{r}_{pp'}\right)  \notag
\end{equation}
and plot the magnitude $|S_M|$ in \figref{fig:vison_3way} (b). We have chosen a columnar background dimerization comprising horizontal background dimers stacked in columns along the $y$-axis, as shown in \figref{fig:orientedSL}; this choice breaks the translational symmetry of the lattice and thus requires doubling the unit cell the $x$-direction. While $|C_M|$ displays no obvious symmetry breaking, there are peaks in $|S_M|$ at $(0,0)$ \& $(0,\pi$). We have plotted the finite-size scaling of these peaks in \figref{fig:bragg_scaling} which demonstrates that these peaks vanish in the thermodynamic limit. We have additionally studied $S_M$ using different background dimerization and find that the location of the finite-size peaks depend on the gauge choice; however in all cases we find that these peaks vanish in the thermodynamic limit. The vanishing of the $(0,\pi)$ peak demonstrates the restoration of rotational and translational symmetries in the thermodynamic limit, and the additional vanishing of the $(0,0)$ peak demonstrates the lack of magnetic vortex condensation, which is a mechanism for destroying topological order and breaking lattice symmetries~\cite{Jalabert1991,Ralko2007,Huh2011}. \figref{fig:vison_cor} shows the decay of $|C_M|$ along one of the axes of the dual lattice which we find decays exponentially with distance, with a correlation length $\chi\approx 3$. These exponentially decaying magnetic string correlations are therefore consistent with a symmetric dimer liquid with no magnetic vortex condensation.

\subsection{Monomer deconfinement}

\begin{figure}[]
        \centering
        \includegraphics[width=1.0\columnwidth]{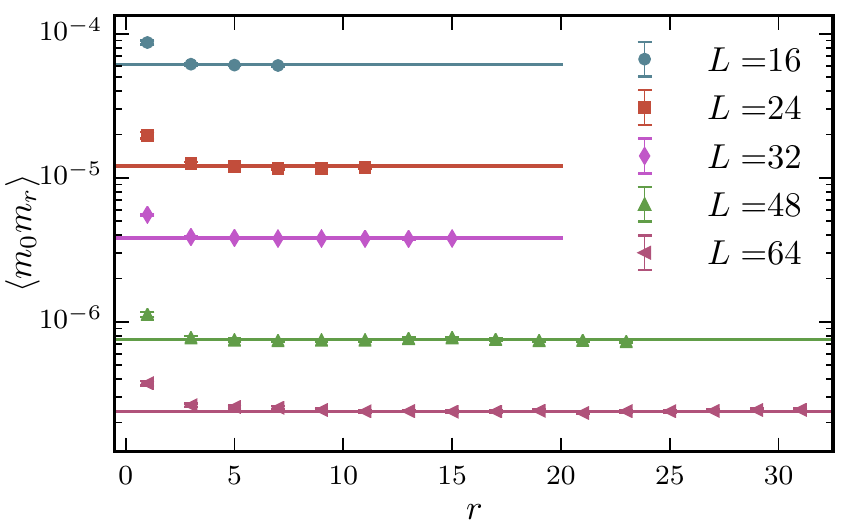}
        \caption{ The monomer-monomer correlation function $\langle m_{\bm{0}} m_{\bm{r}} \rangle$ along a coordinate axis in the QDPM RK ground state doped with a single pair of monomers on lattices up to $L=64$. The horizontal lines correspond to $L^{-4}$ for the system size corresponding to the color.}
        \label{fig:monomer_correlations}
    \end{figure}

To study the confinement of $Z_3$ electrically charged defects in the QDPM, we have Monte Carlo sampled the RK ground state doped with a single pair of monomers. We initialize the system with a monomer on each sub-lattice and supplement the local Monte Carlo updates with updates of the form:
\begin{equation}
\monomermvA  \Longleftrightarrow \monomermvB  \notag
\end{equation}
which appear to provide ergodic dynamics in the monomer-dimer-pentamer configuration space. Notice that monomers remain on the same sub-lattice, which is consistent with their assignment of a $Z_3$ electric charge of $\pm1$ (see \tabref{tab:defects}) and the positive and negative monomers live on different sub-lattices. We then can test for the confinement of monomers by studying the monomer-monomer density correlation function $\langle m_{\bm{0}} m_{\bm{r}} \rangle$, which is plotted in \figref{fig:monomer_correlations}. If the monomers are confined, we expect $\langle m_{\bm{0}} m_{\bm{r}} \rangle$ to decay to zero at large distances; conversely if monomers are deconfined, we expect the monomers to be randomly distributed on the lattice, such that $\langle m_{\bm{0}} m_{\bm{r}} \rangle \sim L^{-4}$ at large separations. \figref{fig:monomer_correlations} demonstrates that $\langle m_{\bm{0}} m_{\bm{r}} \rangle$ decays to $L^{-4}$ for each system size, indicating a deconfinement of monomers. This is in contrast with the QDM where monomer correlations decay as a power-law. Thus we find that $Z_3$ electric charges are deconfined in the QDPM.

\section{Discussion}
\label{sec:discussion}

We have introduced the quantum dimer-pentamer model on the square lattice and studied the ground state at the RK point. Using numerical calculations, we have demonstrated that the RK state of the QDPM is a dimer liquid without spontaneously broken symmetry. Our numerical calculations explicitly show how the extensive topological degeneracy of the square lattice QDM has been reduced to a finite 9-fold degeneracy in the QDPM on a torus. We find that the dimer correlations decay exponentially with distance in contrast to the power-law decay of dimer correlations in the QDM. Additionally, we also have demonstrated that monomer defects are deconfined in the QDPM, in contrast with the critical confinement of monomers in the QDM. Finally, we find that the $Z_3$ magnetic string correlations decay exponentially and display no evidence of $Z_3$ vortex condensation; this is consistent with $Z_3$ vortices being gapped quasiparticles excitations above the QDPM ground state. These results suggest a the low energy physics of the QDPM is described by a $Z_3$ gauge theory with a $Z_3$ topological ordered ground state. Thus we have demonstrated that systems with Ising degrees of freedom and local constraints can lead to $Z_N$ topological order for $N>2$, providing the a new stable phase of matter beyond the commonly seen $Z_2$ topological spin liquid.

Future work could directly probe the existence of a finite gap above the RK point by studying the imaginary time dynamics of correlations functions using Monte Carlo methods similar to those used in here~\cite{Henley1997,Henley2004a}; such an approach could be used to explore if the low energy excitations are best described by $Z_3$ magnetic vorticies~\cite{Ivanov2004,Ralko2007,Misguich2008d}. Additionally, one could conclusively demonstrate the existence of $Z_3$ topological order by computing the topological entanglement entropy in the QDPM ground state using Monte Carlo methods~\cite{Levin2005a,Kitaev2006b,Hastings2010}. 

A finite gap above the RK point suggests that a stable topologically ordered phase should exist surrounding the RK point. While the Monte Carlo techniques used in this manuscript are limited to the RK point, the phase diagram near the RK point can be explored in future work using quantum Monte Carlo methods, as the QDPM has no sign problem. Away from the RK point, the condensation of $Z_3$ magnetic vortices could drive a continuous phase transition out of the topologically order phase, analogous to vison condensation in $Z_2$ spin liquids~\cite{Jalabert1991,Ralko2007,Huh2011,Hao2014,Slagle2014}; if such a continuous transition exists in the QDPM it may provide access to a new universality class of quantum transitions~\cite{XU2012}.

Finally, given the stability of the topologically ordered phase, it may be possible to realize the QDPM in a more microscopically realistic spin model with two body interactions. The local constraints of the QDM can be realized in two-body spin-$1/2$ and Bose-Hubbard models, and the QDM Hamiltonian can arise perturbatively in the constrained low energy Hilbert space~\cite{Balents2002a,Zhitomirsky2005,Isakov2006c,Albuquerque2008}. Thus future work, it may be possible to engineer the QDPM in the low energy subspace of an experimentally realizable two-body model, providing a route to the experimental realization of a $Z_3$ topologically ordered spin liquid. 

\section{Acknowledgements}

We thank T. J. Volkoff for discussions and a careful reading of the manuscript. Computations were performed on the Vermont Advanced Computing Core supported by NASA (NNX-08AO96G).

\appendix

\section{$Z_3$ Gauge theory on the square lattice}
\label{sec:Z3GT}
Here we will review a string-net implementation of $Z_3$ gauge theory on the square
lattice~\cite{Horn1979}. We define a three -dimensional Hilbert space on each link, corresponding to the presence or absence of an oriented string segment that carries $Z_3$ electric flux, which are eigenstates of the Hermitian $Z_3$ electric flux operator $E_{vv'}$
\begin{align}
        & E_{vv'}  \eket{\vvemptylink} = 0
        ,
        \\
        & E_{vv'}  \eket{\rightarrowlink} = + \eket{\rightarrowlink}
        ,
        \\
        & E_{vv'} \eket{\leftarrowlink} = - \eket{\leftarrowlink}
        .
\end{align}

\begin{figure}[t]
    \centering
    \includegraphics[width=1.0\linewidth]{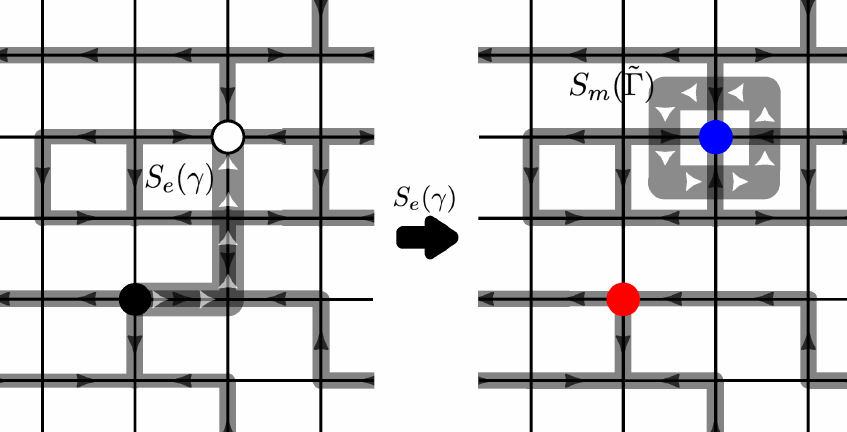}
    \caption{ ({\it left}) An example of a charge free $Z_3$ electric string configuration. The electric string operator $S_\gamma^e$ is represented by the thick gray line, oriented from the white to black vertex. ({\it right}) An example of a $Z_3$ electric string configuration with a $+1$ ($-1$) electric charge at the red (blue) vertex; this string configuration results form the action of $S_\gamma^e$ on the left panel. The magnetic loop operator $S_m (\tilde{\Gamma})$ can detect the presence of the electric charges.  }
    \label{fig:example_elec_string}
\end{figure}

At each vertex we define a local unitary gauge transformation $G_v$,
\begin{equation}
G_v \equiv \prod_{v'} P^\dagger_{vv'}  \notag
\end{equation}
where the product is over all nearest neighbor vertices $v'$, and we have defined the unitary operator
\begin{equation}
P_{vv'}^\dagger \equiv e^{i 2\pi/3 E_{vv'}}.  \notag
\end{equation}
We consider eigenstates of $G_v$ to have a definite $Z_3$ electric charge at vertex $v$. The charge-free sector is the gauge invariant subspace where
\begin{equation}
G_v \eket{\psi} = \eket{\psi} \quad \forall v.  \notag
\end{equation}
This gauge invariant subspace corresponds to states where the ``string flux" vanishes modulo 3 at each vertex, thus including vertices where three inward or three outward strings meet at a vertex. For eigenstates of $G_v$ with  nonzero electric charge
\begin{equation}
G_v \ket{\psi} = e^{\pm 2\pi/3 i} \ket{\psi}  \notag
\end{equation}
which corresponds to the presence of a $Z_3$ charge of $\pm1$ at vertex $v$. A vertex with a positive (negative) charge has a net outward (inward) string flux of 1 modulo 3, or net inward (outward) flux of 2. \figref{fig:example_elec_string} shows examples of charged and charge free string configurations.

To provide dynamics, we define a unitary operator electric flux ladder operator $Q^\dagger_{vv'}$ that cyclically increments the electric flux:
\begin{align}
        & Q^\dagger_{vv'}  \eket{\vvemptylink} = \eket{\rightarrowlink}
        ,
        \\
        & Q^\dagger_{vv'}  \eket{\rightarrowlink} = \eket{\leftarrowlink}
        ,
        \\
        & Q^\dagger_{vv'}  \eket{\leftarrowlink} = \eket{\emptylink}
\end{align}
The simplest gauge invariant Hamiltonian is then
\begin{equation}
H_{Z_3} = - J  \sum_p  \left( Q^\dagger_{v_1^pv_2^p} Q^\dagger_{v_2^pv_3^p} Q^\dagger_{v_3^pv_4^p} Q^\dagger_{v_4^pv_1^p} + h.c. \right)  \notag
\end{equation}
where $v_i^p$ represent the four vertices in the plaquette $p$, ordered cyclically by nearest neighbors.

We can define two oriented open string operators which commute with $H_{Z_3}$ everywhere except at the endpoints. The electric string operator $S_e$, which acts on an open, oriented string on the lattice $\gamma_{vv'}$ with endpoints $v,v'$, and creates a $+1$ ($-1$) charge at vertex $v$ ($v')$, is defined as
\begin{equation}
    S_e \left( \gamma_{vv'} \right) \equiv \prod_{l \in \gamma_{vv'}} Q_{v_l v_l'}^\dagger  \notag
\end{equation}
where the product is over all links $l$ in $\gamma_{vv'}$, $v_l$ and $v_l'$ are the two vertices in $l$, ordered according to the orientation of $\gamma_{vv'}$ (see \figref{fig:example_elec_string}). The magnetic string operator $S_m$ is defined by an open, oriented string on the dual lattice and creates a magnetic vortex-antivortex pair on plaquettes at the endpoints of $\tilde{\gamma}_{pp'}$
\begin{equation}
    S_m \left( \tilde{\gamma}_{pp'} \right) \equiv \prod_{l \perp \tilde{\gamma}_{pp'}} P_{v_l v_l'}^\dagger  \notag
\end{equation}
where the product is over all links $l$ that cross $\tilde{\gamma}$ and $v_l$ ($v_l'$) is the vertex to the left (right) of the string as $\tilde{\gamma}$ is traversed from $p$ to $p'$. 

Note that  $S_e$ and $S_m$ do not commute if their corresponding strings intersect--this underlies the non-trivial statistics of electric charges and magnetic vortices. Consider the magnetic string operator $S_m(\tilde{\Gamma})$ defined by an oriented \emph{closed} loop $\tilde{\Gamma}$ on the dual lattice. This operator corresponds to the creation of a vortex-antivortex pair, winding of the vortex around $\tilde{\Gamma}$, and annihilation of the vortex pair. A state with definite electric charge inside $\tilde{\Gamma}$ will be an eigenstate of $S_m(\tilde{\Gamma})$, with eigenvalue $e^{ i 2 \pi/3 q}$, where $q=0,\pm1$ is the net $Z_3$ electric charge (modulo 3) inside $\tilde{\Gamma}$.  Thus for $\tilde{\Gamma}$ winding around an isolated electric charge, we see that electric charges and magnetic vortices are anyons with relative braiding statistics with a statistical phase of $e^{\pm i 2\pi/3}$.

Additionally, these string operators give rise to a topological degeneracy in systems of non-trivial topology. For example, consider the charge free sector on a torus. We can consider two topologically non-trivial closed loops $\tilde{\Gamma}_{X}$ \& $\tilde{\Gamma}_{Y}$ which have a single winding around the $x$ or $y$ directions of the torus. Since $S_m(\tilde{\Gamma}_{X})$ commutes with $H_{Z_3}$, we can label the energy eigenstates by the eigenvalues of $S_m(\tilde{\Gamma}_{X})$, and similarly for $\tilde{\Gamma}_Y$: $\{1,e^{ i 2\pi/3},e^{- i 2\pi/3}\}$. Thus $H_{Z_3}$ has a 9-fold ground state degeneracy on the torus, and the ground states subspace can be labeled by the $Z_3$ electric flux winding about each axis of the torus.
            
\section{Dimer order parameters QDPM RK ground state}
\label{sec:dimer_order_parameters}

\begin{figure}[t!]
    \centering
    \includegraphics[width=1.0\linewidth]{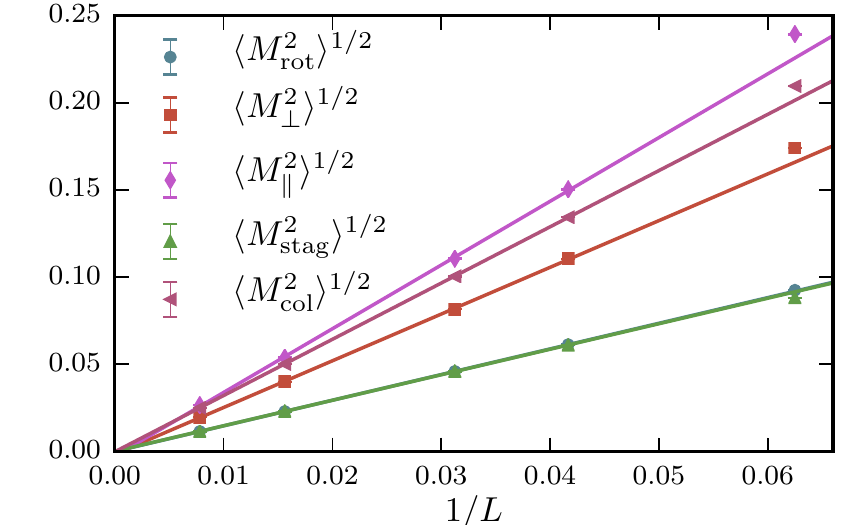}
    \caption{Symmetry breaking order parameters defined in the text computed for the RK point of the QDPM. The lines are fits to $L^{-1}$ scaling. In all cases we find they vanish in the thermodynamic limit, demonstrating the lack of symmetry breaking.}
    \label{fig:order_params}
\end{figure}

To explicitly test for the possibility of translational and rotational symmetry breaking corresponding to the features in $S_d$ at the Brillouin zone edges, we define several order parameters which would capture such potential symmetry breaking. For rotational symmetry breaking we define~\cite{Syljuasen2005} 
\begin{equation}
\Mrot \equiv  \frac{2}{L^2} \left(N_v - N_h\right) \label{eq:Mrot},
\end{equation}
where $N_v$ ($N_h$) is the total number of vertical (horizontal) dimers; note that $\exval{M_{\rm{rot}}^2 } \neq 0$ implies there is rotational symmetry breaking. To capture translational symmetry breaking by one lattice site in a direction perpendicular to the majority of the dimers we define~\cite{Syljuasen2005}
\begin{align}
\Mperp \equiv  \frac{8}{L^2} \sum_{\Box} \Bigl(  &(-1)^{r_x} \theta\left(\Mrot \right) \Pi_\Box^v \notag\\
+ &(-1)^{r_y}\theta\left(-\Mrot \right) \Pi_\Box^h\Bigr) \label{eq:Mperp},
\end{align}
where $\theta(x)$ is the Heaviside function, and $\Pi_\Box^h$ ($\Pi_\Box^v$) is one if the plaquette $\Box$ has two horizontal (vertical) dimers, and zero otherwise. Similarly, we define $\Mpar$ to be given by \Eqref{eq:Mperp} with $r_x$ and $r_y$ swapped between terms. Additionally, we define order parameters that indicate ``columnar" and ``staggered" dimer order~\cite{Syljuasen2005}:
\begin{align}
\MstagSq \equiv  \frac{4}{L^4} \biggl[ &\biggl( \sum_{\ell_h} (-1)^{r_x+r_y} n_d \left(\vec{r}_{\ell_h}\right) \biggr)^2  \notag\\
+&\biggl( \sum_{\ell_v} (-1)^{r_x+r_y} n_d \left(\vec{r}_{\ell_v}\right) \biggr)^2\biggr]\label{eq:Mstag},
\end{align}
\begin{align}
\McolSq \equiv  \frac{4}{L^4} \biggl[  &\biggl( \sum_{\ell_h} (-1)^{r_x} n_d \left(\vec{r}_{\ell_h}\right) \biggr)^2 \notag \\
+ &\biggl( \sum_{\ell_v} (-1)^{r_y} n_d \left(\vec{r}_{\ell_v}\right) \biggr)^2  \biggr]\label{eq:Mcol},
\end{align}
where the sums over $\ell_h$ ($\ell_v$) are taken over all horizontal (vertical) links. In \figref{fig:order_params} we have plotted the finite-size scaling of these order parameters; we find in all cases that while finite size systems display symmetry breaking due to finite-size effects, these order parameters all vanish in the thermodynamic limit, as demonstrated by the extrapolation shown in \figref{fig:order_params}.

\FloatBarrier

\bibliographystyle{apsrev4-1}
\bibliography{QDPM}

\end{document}